\documentclass[twocolumn]{emulateapj}
\submitted{Science with Parkes @ 50 Years Young, 31 Oct. -- 4 Nov., 2011}

\shorttitle{Dishing up the Data: A Decade of Space Missions}
\shortauthors{John Sarkissian}

\begin{document}

\title{Dishing up the Data: A Decade of Space Missions}

\author{John Sarkissian}

\affil{CSIRO Astronomy and Space Science, Australia
  Telescope National Facility, PO Box 276, Parkes, NSW 2870, Australia} 

\email{John.Sarkissian@csiro.au}

\begin{abstract}
The past decade has seen Parkes once again involved in a wide range of
space tracking activities that have added to its illustrious
legacy. This contribution is a personal recollection of those tracking
efforts -– both real and celluloid. We begin in a light-hearted vein
with some behind-the-scenes views of the popular film, ``The DISH'',
and then turn to more serious contributions; discussing the vital role
of the telescope in alleviating the great ``traffic jam'' at Mars in
2003/04 and salvaging the Doppler Wind Experiment as the
Huygens probe descended though the atmosphere of Saturn’s largest
moon, Titan, in mid-decade. We cap off the decade with a discussion of
the search for the missing Apollo 11 slow-scan TV tapes.
\end{abstract}


\section{Introduction}
In 1959, well before the construction of the Parkes telescope was completed, the Jet Propulsion Laboratory (JPL) proposed using Parkes in a cooperative space exploration programme. This involved the occasional use of Parkes, for data acquisition of a short-term nature, where an extremely strong and reliable signal was required. To this day, it is still the rationale for Parkes' support of space missions.

The film, The DISH, was a fictional account of the role played by the Parkes Radio Telescope in tracking and communicating with the Apollo 11 mission. However, the events depicted in the film represent only a single snapshot of the role played by Parkes in space tracking history. This presentation is a personal recollection of the tracking efforts of the past decade at Parkes; both real and celluloid.

\ 
\vskip0.2cm
\section{The dish on The Dish}
In October 1997, John Brooks, the deputy Director of ATNF, received a faxed enquiry, from an author in the UK, requesting the dates that the Parkes Observatory had tracked the various Apollo missions to the Moon. He passed that request onto Marcus Price, the OIC of Parkes Observatory, who had withdrawn the Observatory's log books from the National Archives a year or so earlier. The books were still at the Observatory, so I was asked to look through them and send the author the dates he wanted. As a long time Apollo enthusiast, I had always believed that the Parkes Observatory had received the TV pictures of the historic moonwalk as Neil Armstrong stepped onto the Moon at 12:56 PM on Monday, 21 July 1969. So, you can imagine my surprise when, looking through the log books, I discovered that the Parkes telescope was not tracking the Moon until 1:02 PM – a full six minutes after Armstrong had stepped onto the Moon. I was intrigued by this discrepancy and was determined to understand why. So began what eventually became my decade and a half investigation into the Parkes Observatory's role in support of the Apollo 11 mission.

Almost a year later in October 1998, John Reynolds, the newly appointed acting OIC of Parkes Observatory, received an enquiry from the Working Dog production company that they were interested in doing a story on the Observatory's role in the Apollo 11 mission. They wanted to visit the site later that week for a quick reconnoiter. They were interested to know if the telescope still looked the way it did in 1969. We all assumed that they were planning a documentary; after all, what else can you do with Apollo 11 and Parkes. John was due to be away on a management course on the day of their visit, so as the resident Apollo ``expert'', he asked me to look after them and answer any Apollo related questions. They arrived around noon on Friday, 23 October just as we were busy preparing for the Observatory Open Days which were due to start the following day. Things were pretty hectic on site, but we managed to get them up onto the dish surface and answer all their questions such as how strong were the signals from the Moon, how large an antenna was required to detect them and whether the signals could be detected with backyard telescopes etc. They also discussed amongst themselves the best places to film characters with names like Mitch, Al and Cliff. Since I knew there weren't any Mitchs or Als at Parkes in 1969, it slowly began to dawn on me that they weren't planning a documentary at all, but rather, a feature film. Even more bizarrely, I suspected that it might actually be a comedy.  Over the next six months, events moved quickly and contract details were drawn up.

Principal photography at the Observatory began on Monday, 24 May and ended on Friday, 4 June 1999. The first scenes shot were the helicopter flyovers. As we drove into work that morning on the Observatory bus, I remember seeing John Reynolds standing in the field near the observer's quarters, looking very concerned. The agreement with the helicopter crew had been that for safety reasons, they could get no closer than 300 metres to the dish. Unfortunately, the crew took that to mean that they could get extremely close of the dish. The sight of the helicopter rotors just a few metres from the edge of the dish surface, as they filmed Brett Preisig descending the focus cabin ladder leg, made John very nervous.  He was greatly reassured though, when he learned that the pilot was Greg Duncombe, who had flown the helicopter for ``The Matrix'' a year earlier. 

About a month prior to principal photography, I was asked by the
production company to provide some figures for a blackboard
calculation that was to appear in the film. The producers described a
scene where the Parkes astronomers try and calculate the position of
the Apollo 11 spacecraft after they ``lose'' it earlier in the film. The
producers were after any old figures that would look convincing. I was
hesitant. I'd heard a story a few years earlier relating to the
Hollywood film, ``Good Will Hunting''. The film featured a mathematics
genius who could solve complex problems posed by a mathematics
professor played by Robin Williams. I'd heard that soon after the film
was released on video, a real mathematics professor in the United
States, hired the video and paused it at the scenes showing the
calculations on a blackboard. He wrote an article ridiculing the
mathematics as simplistic gibberish. I wasn't sure how true this story
was, but I was determined not to suffer the same fate. So, I insisted
on doing it correctly. I acquired from colleagues in the United States
a book titled, ``Apollo by the numbers''. This contained many useful
facts including three of the six orbital elements of Apollo 11. I used
these and made educated guesses of the other three elements to
calculate the trans-lunar trajectory of Apollo 11. I compared my
calculations with the actual Parkes logbook entries and found them to
agree very closely. I wrote a fifteen-page report explaining the
method along with a sheet showing the blackboard calculations.  In my
report, I described how calculating a spacecraft orbit was a complex
n-body problem, but a good approximate solution could be found if a
two-body problem was assumed when the spacecraft was still close to
the Earth. I described how the astronomers could use the calculated
position as a reference point and then perform a raster scan about
this point to find the spacecraft. To my surprise, I discovered when
the film premiered the following year, that the NASA character, Al,
used these very words to suggest a solution to the problem of
``finding'' Apollo 11. I was disappointed therefore, that despite all my
efforts, the film characters ignored the calculations entirely. My one
consolation is that no one can possibly fault the calculations.

The film premiered on 19 October 2000, just a few weeks after the Sydney Olympics. It was a huge success, being the largest grossing Australian film for that year. The reviews of the film were universally positive. The film, re-introduced the Parkes Telescope to a new generation of Australians and, most importantly, to a new generation of politicians. The visitor numbers jumped from about 50,000 a year to over 120,000 in the year following the premier. The visitor numbers continued to rise, peaking two years later at 138,000. Today, they are still around 100,000 a year - about double the pre-film numbers. Fortunately a long-awaited upgrade and expansion of the Visitors Centre was completed around this time, which has
greatly contributed to the sustained visitor numbers. 

\section{The Traffic Jam at Mars – 2003-04}
In 2003, Parkes was approached to support NASA's Asset Contention Period (ACP) from early November 2003 to mid-February 2004. This ACP was more popularly termed ``the traffic jam at Mars''. The congestion at Mars included the Mars Global Surveyor and Mars Odyssey 2001 satellites already in orbit. They were to be joined by the two Mars Exploration Rovers (MER), Spirit and Opportunity, plus the European Mars Express satellite which was to deploy the British Beagle 2 lander shortly after reaching orbit on Christmas day 2003. Finally, the Japanese probe, Nozomi (meaning hope), was to orbit Mars in January 2004.

In addition, there were many other spacecraft strewn across the solar system requiring tracking support. These included Stardust, Deep Impact, Genesis, Cassini, Voyager 2 and SERTF (IR Telescope). Many of these were close to the same celestial longitude as Mars. This crowding of spacecraft in one part of the sky, posed a scheduling dilemma for the mission planners. Normally, the spacecraft are more-or-less distributed uniformly throughout the solar system, so that the three tracking stations of NASA's Deep Space Network (DSN), can easily handle the workload. In 2003, there were so many spacecraft clustered about Mars, that scheduling adequate coverage for the various missions became a major concern. A sophisticated scheduling system with a team of hundreds of negotiators around the world ensured that each mission's priorities were met. In order to augment the capabilities of the Canberra Station at Tidbinbilla, and to provide greater flexibility for the schedulers, Parkes was contracted to act as an extra receiving antenna. This freed Tidbinbilla to concentrate on those spacecraft requiring two-way communication. Within the DSN, Parkes was identified as DSS-49 (Deep Space Station 49).

The \$3 million contract with NASA involved resurfacing the dish, providing a new X-band receiver and performing the daily tracks over a 4-month period. Preparations for these tracks began two years earlier. In order to enhance the efficiency of the telescope at these frequencies, in March 2003, the dish surface was upgraded. The resurfacing extended the perforated aluminium panels out to 55 metres diameter. The panels were manufactured by the firm, Sydney Engineering Sales (SES), to ATNF specifications using the methods first developed by Barry Parsons and Don Yabsley in the 1980s. Barry Parsons was the ATNF Project Engineer and Mike Kesteven was the Project Scientist for the upgrade. Tom Lees supported the contractor during the resurfacing on site. Ken Skinner was the SES project engineer who supervised the manufacture of the panels and their installation.

Following the installation of the panels, holography measurements showed that the surface accuracy of the dish was 0.8 mm RMS for the inner 55 metres.

The spacecraft at Mars would be transmitting at frequencies close to 8.4 GHz. In order to receive these signals, a new low-noise, X-band receiver was built by the receiver group at Epping led by Graham Moorey and Russell Bolton. The receiver had a 50 MHz wide bandpass centred on 8.4 GHz. It was installed on the telescope in August 2003, and tests showed that it had a Tsys of about 25 K - significantly better than the contractual specs. The combined receiver and dish surface upgrades contributed to making the telescope some 6 dB more sensitive at X-band, that is, a four-fold increase in sensitivity at this frequency - a great boon for radio astronomers.

The official launch of the Parkes Mars Tracks was held on Friday, 31 October. The US Ambassador to Australia, His Excellency, J. Thomas Shieffer, was the guest of honor.  He was accompanied by Mr John Cobb MP, the local member of Federal Parliament, and by Dr Neal Newman, the NASA Representative in Australia. His Excellency was welcomed to the Observatory by Dr John Reynolds, the Officer-in-Charge, who in turn introduced Dr Brian Boyle, the Director of the CSIRO's Australia Telescope National Facility. Present also, was Prof. Brian Schmidt, the then chairman of the AT Time Assignment Committee.

As so often happens during crucial moments in its history, the telescope was subjected to a buffeting from Mother Nature. In a case of life imitating art, high winds and a light rain greeted the visiting dignitaries, but everyone present took it with good humor.

During the formal part of the ceremony, Dr Warren King, Chairman of the CSIRO's IT Manufacturing and Services, representing the CSIRO Chief Executive, Dr Geoff Garrett, outlined the history of Parkes' contributions to space exploration and the long collaboration between NASA and the CSIRO. He described how Parkes was involved in the Mariner 2 and 4 missions that began interplanetary explorations of Venus and Mars in the 1960s. The Apollo manned lunar landing missions were a triumph as were the Voyager 2 encounters of Uranus and Neptune in the 1980's. Finally, he described how in the year-long Galileo tracking support from 1996 to 1997, the observatory maintained a phenomenal 97.25\% uptime and that no tracking time was lost due to operator error. Dr King also described how the dish was the model and inspiration for the large antennas of NASA's Deep Space Network (DSN).

The Ambassador, in his response, described the Parkes Telescope as like a trusted friend, "always there when we need a hand". He said that the relationship between the CSIRO and NASA was very much like that between the United States and Australia, as friends that share common values and dreams. The Ambassador also described how he was first asked by President George W. Bush to be the US Ambassador in April 2001; "I went back to Dallas that weekend and of all things that were playing that week at the theatre was "The DISH". My wife and I went to see it and I knew this (being Ambassador to Australia) was going to be a lot of fun", he said.

To commemorate the upgrade, and officially launch the Mars Tracks, three spare panels were set up at the base of the tower. The dish was lowered to within 2 metres of them. At this moment, right on cue, the winds suddenly picked up, peaking at over 80 km/h. A bottle of champagne was attached to the edge of the dish by a length of rope. Accompanied by Dr John Reynolds and Mr John Cobb MP, the Ambassador pulled on the rope, which raised the bottle, and to a chorus of cheers, sent it smashing into the centre panel.  In an allusion to the film, The DISH, Dr Warren King then presented the Ambassador with a classic Australian Cricket bat. All the Parkes Staff signed the front of the bat, and all the ATNF staff from our Sydney headquarters signed the back. In a reciprocal gesture, the Ambassador presented Dr John Reynolds with a Grand Slam Louisville Slugger baseball bat. 

The Ambassador, along with the media, was then treated to a hayride. Up on the surface, a quick spot of cricket was played. With John Reynolds bowling and Neal Newman keeping wicket, the Ambassador eyed the ball and hit it for six, sending it almost into orbit - Bradman would have been proud!  Pictures of this cricket match were splashed across the national media in that evening's TV news bulletins and the next day's newspapers.

Coming back down to Earth, the Ambassador was escorted through the tower to inspect the control room. Dr Stacy Mader, who had been controlling the dish during the ceremony, showed the Ambassador the Manual Control Panel and invited the Ambassador to take control and stow the dish, which he was delighted to do. The Ambassador was then introduced to the visiting astronomers in the new control room. Dr Richard Manchester and Dr Michael Kesteven explained their research and introduced their students, Dr Nina Wang, Aidan Hotan and Aris Karastergiou. 

The media then had opportunities to interview the Ambassador. Following the press interviews, our guests gathered in the dish café for lunch over music and drinks. Following lunch, the Ambassador presented the Observatory with a framed Louisville Slugger Major League baseball kit. The ball was signed by the Ambassador, and was put on display in the Visitors Centre to commemorate the visit. We understand that the bat is considered to be the Rolls Royce of baseball bats. The inscription on the frame reads: `Presented to CSIRO and the Parkes Observatory - big hitters on the NASA team'. 

31 October, was also the 42nd anniversary of the Parkes Telescope's official commissioning. Mr John Cobb MP, praised the staff of the Observatory for their efforts in maintaining the telescope at the forefront of scientific research and for preparing it for the tracks. He toasted the continued success of the observatory, something that all Parkes residents and Australians in general were immensely proud of.  Then, lead by John Reynolds in a rousing chorus of Happy Birthday, the Ambassador and John Cobb cut the dish's birthday cake.  Despite the weather, the occasion was marked by good humour and good will.

On a personal level, I was introduced to the Deputy Head of the US Mission, William Stanten. During our conversation, we realised that both our grandmothers were born in the same city in Armenia. We optimistically concluded that we were very likely distantly related. After almost a century, two, long-lost cousins had found each other.

The formal tracking period commenced on Monday, 3 November 2003 and continued until Sunday, 22 February 2004. Stacy Mader and John Sarkissian conducted the tracks over this four-month period, during which Parkes was given the task of mainly tracking Voyager 2 and Mars Global Surveyor (MGS).  The Voyager 2 tracks were especially important since it was approaching the Solar System's termination shock. In October 2003, Voyager 2 detected a shock wave from solar storms. Powerful flares had hurled billion-ton clouds of gas across the Solar System. Over the following weeks, Voyager 2 measured their speed, composition, temperature and magnetism.

Two days after the end of the formal tracks, on Tuesday, 24 February, Dr Charles Elachi, the Director of JPL, visited the Observatory and personally thanked the staff for all their efforts in support of the ACP.

In July 1965, the Parkes Observatory helped track the Mariner 4 spacecraft as it flew past Mars, returning the first-ever close-up images of the Martian surface. The ACP tracks continued the Parkes telescope's proud legacy of exploring the red planet.

\section{Huygens Heroes – 14 January 2005}

The NASA/ESA Cassini-Huygens mission was launched on 15 October 1997 and spent 7 years cruising to Saturn. It arrived on 1 July 2004 and went into orbit about the planet.

The spacecraft was in two parts: The Cassini spacecraft was an orbiter built by NASA and is still in orbit studying the planet, its rings and retinue of Moons. The Huygens probe was built by the European Space Agency (ESA) and was intended to study the atmosphere of Titan.

In the original mission plan, the Huygens probe was scheduled for deployment soon after it arrived at Saturn in November 2004. However, in 2000, during the long cruise to Saturn, inflight tests by ESA engineer Boris Smeds, found that the Huygens receiver onboard Cassini could not track the Doppler shift of the Huygens signal and decode the data stream correctly.  The hardware of Cassini's receiver was designed to be able to receive over a range of shifted frequencies. However, the firmware failed to take into account that the Doppler shift would have changed not only the carrier frequency, but also the timing of the payload bits, coded by phase-shift keying at 8192 bits-per-second.

The solution to this design flaw was to change the landing from November 2004 to January 2005. With the new trajectory, the Huygens probe would descend at a more perpendicular angle relative to Cassini, thus reducing the Doppler shift and enabling the receivers to correctly receive and decode the Huygens transmissions.

During the descent, the Cassini spacecraft would point towards Huygens, receiving its data and recording it onto its solid state recorders. Data from the probe would not be received at Earth until after the Cassini spacecraft had moved out of range of the Huygens probe, which by then would be resting on the surface. The Cassini spacecraft would then point back toward the Earth and transmit the data. The Deep Space Network's (DSN) tracking station at Tidbinbilla near Canberra was to be the first to receive this information.

The Huygens probe was designed to transmit its data to Cassini in two separate channels, one at 2040 MHz and the other at 2090 MHz. This data was transmitted alternately by two antennas 6 seconds apart, that is, data would first transmit in channel A by one antenna then, 6 seconds later, the next item of data would transmit in channel B by the other antenna and so on. This redundancy was to allow the signal reception in case the probe was swinging like a pendulum in the wind. By buffering the data, and spacing the transmissions, some data reception was guaranteed as the antennas alternately swung in and out of view of Cassini.

The signal from the Huygens probe was never intended to be detected on Earth. However, with the new descent profile, it was realised by astronomers that the antennas onboard Huygens would be pointing only 30 degrees from Earth during the descent through Titan's atmosphere. The signal strength from the probe would be strong enough to allow the direct detection of the signal from Earth. The signal was too weak to receive any data at Earth, but it was strong enough to act as a beacon. This meant that if a network of radio telescopes could be setup, the signal could be used to pinpoint the probe's position to within one kilometre as it descended to the surface of Titan. This would be a phenomenal achievement, especially considering that the probe would be 1.2 billion kilometres from Earth at the time.

In addition, during the descent, the frequency of the signal would be Doppler shifted a tiny but measurable amount as the probe was blown along by the winds of Titan. By combining the measurements of the Doppler shift with the position of the probe, a three dimensional velocity profile of the winds in Titan's atmosphere could be determined.

Since the descent of the probe would occur in view of the Pacific Ocean, in February 2004, Leonid Gurvits of the Joint Institute for VLBI in Europe (JIVE) organised a VLBI network around the Pacific Rim to take advantage of this fortuitous alignment. The plan was to track the weak channel A (2040 MHz) signal and pin-point the position of Huygens to within just one kilometre and to obtain the transverse velocity (on the plane of the sky) as well as the Doppler (radial) velocity.

JIVE arranged for up to 17 antennas to be linked together in a VLBI network. Five of these antennas were located in Australia. They included the CSIRO's Parkes, Narrabri and Mopra Observatories, and the University of Tasmasnia's antennas in Hobart, Tasmania and Ceduna, South Australia. 

Other antennas were located in the United States including the 100-metre telescope at Green Bank, West Virginia, and the antennas of the Very Long Baseline Array (VLBA) at Kitt Peak, Owens Valley, Brewster and Mauna Kea. Two other telescopes in China and Japan were also used.

The Green Bank and Parkes telescopes were essential for the observation because of their large collecting areas. Dr Chris Phillips, of the CSIRO's Australia Telescope National Facility (ATNF), was responsible for co-ordinating the entire Australian effort.

Working under the direction of Dr John Reynolds, Parkes staffers Brett Dawson, John Crocker, Ken Reeves and Jeffrey Vera along with George Graves from the ATNF receiver group, set to work modifying the Observatory's Galileo receiver. In 1996-97, this receiver was used by the Observatory to receive the signals from the Galileo spacecraft in its yearlong tracking support. They were able to match the S-band SETI feed from the Observatory's Project Phoenix (SETI) receiver, to the input aperture of the Galileo receiver to allow Parkes to receive the signals from Huygens.

In October 2004, Bob Preston at NASA/JPL, contacted Dr John Reynolds and asked if they could piggyback a Doppler Wind Experiment (DWE) on the Parkes observations. JPL would install a Radio Science Receiver (RSR) in the Parkes control room to enable them to detect the Doppler data in real-time, rather than to wait weeks or months to get it from the VLBI data. The JPL DWE measurements would enable the determination of Titan's wind speed in the Earth-probe direction as a function of altitude. By combining this with the simultaneous Doppler observations in the Cassini-probe direction, the actual direction and strength of the zonal winds of Titan's atmosphere could be determined. The wind-induced motion of the probe could thus be measured to a precision of better that 1 m/s.

On 4 January, a Radio Science Receiver (RSR) from the DSN tracking station at Tidbinbilla was sent to Parkes for use in the DWE. Dr John Reynolds and Barry Turner, installed the receiver in the Parkes control room using the cabling still in place from the NASA Mars tracking operations of a year earlier. The DWE would only be conducted at Green Bank, the VLBA and Parkes. However, the real-time detection of the signal was only possible from Green Bank and Parkes - the VLBA data would be processed much later. 

On 10 January, two JPL engineers arrived from the United States to conduct the DWE. They were Dr Jim Border and Doug Johnston. Dr Border was from the Tracking Systems and Applications Section of JPL with expertise in spacecraft signal measurement. He was at Parkes to make a recording of the Huygens signal using a PC based data recording system. In 1995, he had worked on a similar recording of the Galileo probe as it plunged into Jupiter's atmosphere using the CSIRO's Australia Telescope Compact Array at Narrabri. Doug Johnston was from the Radio Science Group at JPL. He was at Parkes to operate the RSR and to process and analyze the data in real-time. Six months earlier, on 1 July, he had helped to put the Cassini spacecraft in orbit about Saturn.

The first part of the descent of the Huygens probe would be visible from Green Bank and the second part from Parkes. Twenty minutes before the probe was scheduled to land on Titan, it would set at Green Bank. But at Parkes, the expected landing time was just one minute after it was scheduled to rise, that is, the critical events at Parkes would occur right on the telescope's horizon. Since there was about a three minute uncertainty in the expected time of the landing, Dr John Reynolds and Andrew Hunt modified the SERVO computer code to lower the telescope's software-limited, elevation horizon to 30.25 degrees from its usual 30.5 degrees. This extra 0.25 degrees gave the telescope about 3 minutes more tracking time, enough to catch the landing if it occurred earlier than expected. With this new horizon limit, the probe was due to rise at Parkes at 10:29 PM (AEST) with the nominal landing time expected at 10:33 PM (AEST). The batteries were expected to last a further one-hour before being exhausted and the transmissions ceased.

The day of the track was a typical January day at Parkes; it was extremely hot and dry with temperatures hovering around 40° C all day. In the late afternoon a summer storm broke over the dish bringing with it a welcome cool change. The winds kicked up a minor dust storm in the parched farm fields surrounding the telescope, but as dusk approached, the winds abated, the skies cleared and a rainbow appeared above the dish. 

Inside the control room the team set about preparing for the track. Parkes Operations Scientist, Dion Lewis, checked and double checked the VLBI recording equipment and cleared lots of disk space for the expected flood of data. Dr Jim Border and Doug Johnston set up their equipment and established a continuous telephone link to JPL's mission control in Pasadena. Dr John Reynolds performed numerous focus and pointing checks of the receiver, fine tuning the system to get the strongest possible signal.

When Huygens entered the atmosphere of Titan, the giant Green Bank telescope in West Virginia, USA, was poised to detect the signal when the transmitters sprang to life. Sure enough, right on schedule at 8:18 PM (AEST), JPL's Dr Sami Asmar, who was based at Green Bank, reported his detection of the signal. A quiet cheer went up in the Parkes control room - we knew we had a mission.

Both Green Bank and Parkes were equipped with spectrum analysers that allowed them to see the signal as a small spike in the pass-band of the receiver. It was this spike that Dr Asmar at Green Bank had reported seeing. At Parkes however, Doug Johnston had the additional capability to further process the data and produce plots of the carrier's Doppler shift variations. The RSRs at Green Bank and Parkes were capable of measuring the Doppler shift of the signal in real-time. Doug could compare these to the predicted Doppler shifts based on a smooth atmospheric descent model. Any variation from these predicts was an indication of what the real atmosphere of Titan was doing, that is, it was a measure of the winds on Titan. Within just three minutes of the initial signal reception, Doug had transferred the Green Bank data over the JPL network and processed it at Parkes. The difference between the real and predicted Doppler shifts was plotted.  It showed that the probe was deviating significantly from the expected descent profile. At first the detected signal was just a little off the predicts, but as it descended further, the deviations increased and fluctuated. The winds on Titan were furiously blowing the little probe about. It was an amazing feeling to realise that we were the first people ever to ``see'' the winds of Titan.

At 8:32 PM (AEST) Doug reported seeing the glitch in the Doppler shift that indicated the main parachute had deployed, abruptly slowing the probe. The parachute had deployed on schedule at 10:32:27 UTC. He immediately notified JPL.

Meanwhile, the tension at Parkes was quietly rising as the time approached for Parkes to take over. One hour before the beginning of the track, Dr John Reynolds switched over to the generator to prevent an unforeseen power loss from disrupting the track. He continued to check the focus and pointing of the receiver. At 10:12 PM (AEST) he slewed the dish over to the horizon waiting for Titan to rise. At the same time the probe set at Green Bank.

As the minutes passed, we all gathered in front of Doug Johnston's console waiting for the beginning of the track and confirmation that we were receiving the signal. Right on schedule at 10:29 PM (AEST) the dish began tracking and the signal appeared on the spectrum analyser. Doug announced we had the signal and we all cheered and began clapping with excitement. The signal was 2.5 times stronger than expected. We were elated. Dion Lewis confirmed that the VLBI data taking had begun. We were getting good VLBI data, so we knew that the primary part of our support would succeed.

Four minutes later, the nominal landing time had passed and the probe still hadn't landed. We knew that there was up to a 3-minute uncertainty in the landing time so we weren't concerned. As long as we were getting a strong signal, all was fine. As time passed, the engineers at the European Space Operations Centre (ESOC) in Darmstadt, Germany, were eager to know if the probe had landed. The tension at ESOC was rising, but at Parkes things were calm since we were still getting a good signal.

At around 10:45 PM (AEST) Doug reported a slight glitch in the Doppler shift but that it was much less than expected for the landing. He had been viewing the plot of the Doppler shift variations, which only showed glitches when they departed from the predicts. Dr Jim Border decided to plot the sky frequency instead, that is, the actual frequency received. Sure enough, there was a large glitch at the suspected landing time of 10:45 PM (AEST). This confirmed the landing. Shortly after 11:00 PM (AEST) Jim and Doug alerted JPL and ESOC and the word quickly spread around the world that the Huygens probe had landed. Cheering erupted in the Parkes control room and congratulations were exchanged. The landing had been a much softer touchdown than expected and it occurred sometime between 10:45 and 10:46 PM (AEST) - 12 to 13 minutes later than expected. It was a second moon landing for Parkes.

Meanwhile, the VLBI observations continued. Every five minutes the telescope would slew away from Huygens and point toward a nearby calibration source for one minute before returning again to Huygens. This is referred to as ``nodding'' and the dish was doing this throughout the track. This is not an ideal situation for tracking spacecraft, but since the DWE was piggybacking onto the JIVE VLBI observation, it was accepted. The landing had occurred during one of these one-minute nods, hence the slight uncertainty in the landing time.

The Huygens batteries were designed to last for only about one hour after the landing. But, as the track continued, the signal remained strong and showed no indication of weakening. As the one-hour mark passed, followed by the two-hour mark, it was obvious that the batteries would last much longer than expected. JIVE astronomers began to contemplate the possibility that the signal may last long enough for it to be detected from Europe when it rose there several hours later. Some last minute, frantic activity was initiated to try and get European antennas up and tracking in time to detect the signals from Huygens. However, because of the non-standard observing frequency and the non-availability of the disk based recording equipment at the various observatories, it was not possible to organise this in time.

Just after mid-night on 15 January (AEST), the DSN's tracking station at Tidbinbilla near Canberra, made contact with Cassini after it had over flown the probe's landing site and turned to face the Earth. Shortly there after, Tidbinbilla began receiving the Huygens data being played back by the mother craft.

The Parkes dish continued to perform flawlessly throughout the track until at 1:56 AM (AEST) Huygens finally set at Parkes, still transmitting strongly. The champagne was duly popped open in celebration. For this moon landing, the high winds were thankfully on Titan and not at Parkes. It had been a magical night!

Later that morning, Dion Lewis removed the disks containing the VLBI data and together with Dr Chris Phillips, who had the data from Narrabri and Coonabarabran with him, they transported them to the ATNF Sydney headquarters on a specially chartered flight. In Sydney, the data was sent along a high-speed data link to the JIVE headquarters at Dwingaloo in The Netherlands. There the Parkes data was processed within the day, confirming that the VLBI experiment was a resounding success. The high-speed link to JIVE had been specially arranged for the purpose, and really kick-started eVLBI in Australia. The need for the chartered flight to Sydney highlighted the lack of dedicated high-speed data links to the Observatories and gave great impetus to rectify this deficiency. Little more than 12 months later we had a dedicated 1Gb/s network link to all three Observatories. 

Shortly after the end of the track, it became apparent that the data being played back by Cassini was missing the 2040 MHz, channel A, telemetry. Apparently, a sequencing error by the ESA controllers in Germany, had resulted in the 2040 MHz receiver on board Cassini not being switched on. This meant that although the data from channel A was transmitted by Huygens it was not received by Cassini. Fortunately, the loss of data from this error was not critical since the data was redundant. None-the-less, only half of the information transmitted was received, including the images. Sadly also, for the DWE, it meant that Cassini could not measure the Huygens Doppler shifts. Since it was this beacon that Parkes was tracking, the Parkes and Green Bank recordings of the channel A signal assumed greater significance in the DWE. In fact, the Parkes and Green Bank observations salvaged the entire experiment - ``sometimes it pays to eavesdrop'' declared Dr Sami Asmar of JPL.

On 9 February, ESA and NASA simultaneously released the results of the Parkes and Green Bank DWE observations. They indicated that the winds on Titan moved in the direction of the Moon's rotation and that the maximum wind speeds were measured to be about 430 km/h at an altitude of 120 km. The winds were weak near the surface, but above 60 km altitude the winds fluctuated greatly indicating significant vertical wind shear. The probe landed with a velocity of only 2.9 m/s, much softer than the expected 5 to 6 m/s.

The Parkes Huygens track was a great team effort and a tribute to the intense scientific collaboration between CSIRO's ATNF, NASA/JPL, ESA and JIVE. It bodes well for future co-operative ventures. 

The excitement experienced at Parkes was repeated at all the radio observatories involved. Together, several major objectives were achieved by the observations:
\begin{itemize}
\item{Highly accurate VLBI measurements of the probes trajectory in the atmosphere of Titan were achieved.}
\item{By combining the VLBI and DWE data a 3-dimensional model of the probe's motion through Titan's atmosphere was reconstructed, essentially mapping the winds of Titan.}
\item{The VLBI tracking increased the scientific return of the mission.}
\item{Radio contact was maintained with the probe after the link with the Cassini orbiter was broken.}
\item{The tracking potential of current and future VLBI instruments for planetary and deep space mission support was demonstrated.}
\item{The Parkes and Green Bank observations salvaged the DWE from total failure.}
\end{itemize}

Postscript: The ESA engineer, Boris Smeds, who identified the design flaw on the Cassini receiver, and which ultimately allowed Parkes to be involved in the mission, was stationed at Parkes in March 1986 as the engineering operations supervisor for the ESA Giotto encounter with Halley's Comet.

\section{The Search for the Missing Apollo 11 SSTV Tapes}

In July 1969, six hundred million people, one-sixth of mankind at the time, witnessed the historic Apollo 11 moonwalk live on television.

At the time, three tracking stations were receiving the TV signals from the Moon. They were the CSIRO Parkes Radio Telescope and the NASA tracking stations at Honeysuckle Creek near Canberra and Goldstone in California. As the signals were being received they were recorded onto magnetic data tapes.

The TV signal that was transmitted directly from the Moon was of a non-standard format that had to be scan-converted on Earth before it could be broadcast worldwide. The scan-conversion process unavoidably produced a lower quality version of the TV pictures. It was this version that was witnessed by the world and archived by NASA. 

Beginning in the late 1990s, I began searching for the original data tape recordings with the aim of recovering the higher quality TV. I was later joined by four colleagues including Stan Lebar, Dick Nafzger, Bill Wood and fellow Australian, Colin Mackellar, In early 2006, I visited the United States and compiled a report on the rationale for the search and the results to date. The report was posted on the Parkes Observatory web site in May 2006. Within a few weeks, interest began to rise, culminating in a Sydney Morning Herald article on 5 August, with the provocative headline, ``One giant blunder for mankind: how NASA lost moon pictures''. This caused a stir with the story going viral on the internet and news reports appearing on the American TV networks and other news organizations worldwide. Interest became so intense that on 16 August the NASA Administrator, Michael Griffin, formalised the search and appointed the NASA Goddard Space Flight Center Deputy Director, Dorothy Perkins, to head the search. Following an exhaustive three-year search, our Apollo 11 Tape Search Team reluctantly concluded in 2009 that the data tape recordings no longer existed. It's believed that the tapes were very likely erased and re-used in the late 1970s and early 1980s as a cost-cutting exercise by NASA. The results of the search were announced by NASA on 16 July 2009, at a press conference in Washington DC, during the 40th anniversary commemorations of Apollo 11.

Fortunately, during the search, many of the very best scan-converted recordings were discovered in various archives around the world. These were digitised and the best parts of each were used to compile a complete recording of the Apollo 11 moonwalk.

The bulk of the footage was sourced from the signals received by the CSIRO Parkes Radio Telescope, but it also included new, never-before-seen footage sourced from the Honeysuckle Creek tracking station.

In 2009, NASA contracted the California company, Lowry Digital (a pioneer in video enhancement), to process and restore this recording for the Apollo 11 40th Anniversary. The restoration involved digitally repairing damaged sections of the recordings, removing noise from the video, correcting for vignetting, stabilising and brightening the TV picture and other adjustments.  The result is the best and most complete video record yet of the Apollo 11 moonwalk.

As Neil Armstrong has commented, ``the restored video is a valuable contribution to space exploration and space communication history''.

The NASA Goddard Space Flight Center produced three archival sets of hard drives containing the complete Apollo 11 restored video – one set was sent to the US National Archives in Washington and another went to the Johnson Space Center in Houston, Texas. The third set was destined for Australia in recognition of the substantial involvement of the Australian tracking stations in receiving the Apollo 11 television signal. They arrived in Australia on 16 August 2011 and delivered to the Canberra Deep Space Communication Complex at Tidbinbilla, having been previously organised by the former Director, Dr Miriam Baltuck.

A week later, on 24 August, Neil Armstrong was in Sydney to address the CPA Australia 125th anniversary celebrations. During his address, Neil Armstrong paid a glowing tribute to the many Australians who worked at the tracking stations and helped to ensure the success of his Apollo 11 mission. Some of them were present in the audience and were individually acknowledged by him.

In a brief ceremony following the event, Neil Armstrong symbolically handed over the Australian disks to Dr Phil Diamond, the Director of the CSIRO Astronomy and Space Science (CASS) - the custodian of the disks in Australia. This ceremony effectively brought the restoration effort to a close.

The Australian disks will eventually be deposited in permanent archival storage, most likely with the National Film and Sound Archive in Canberra.

DVD's of the restored Apollo 11 video were produced and made available for sale in the Observatory Visitors Centre in time for the Parkes Observatory's 50th Anniversary celebrations.

\section{Conclusion}

In the 16 December 2000 edition of the Sydney Morning Herald, Sandra Hall, in a review of the film, The DISH, titled, ``In Peter Pan Country'', wrote: ``The Australia Telescope and its astronomers are among the country's overlooked treasures.''

Today we celebrate the 50th anniversary of the dish, an icon of Australian science. The work performed here has shaped Australian attitudes to science, especially astronomy. It is arguably the reason why astronomy continues to be supported and held in such high regard by both the government and the public at large. With its constant upgrades, the Parkes telescope is assured a long future as it continues to dish up the data for decades to come.

\end{document}